\documentclass[aps,preprint]{revtex4}
\usepackage{hyperref}
\usepackage{amsfonts}
\usepackage{amsmath}
\DeclareMathOperator{\diag}{diag}
\usepackage{amssymb}
\usepackage{graphicx}%
\usepackage{bigints}
\usepackage{graphicx}
\usepackage{subfigure}
\usepackage{epsf}
\usepackage{bm}
\usepackage{amsmath}
\usepackage{amsfonts}
\usepackage{amssymb}
\usepackage{graphicx}
\usepackage{tabularx}
\usepackage{color}%
\providecommand{\U}[1]{\protect\rule{.1in}{.1in}}

\newcommand{\be}{\begin{equation}}
\newcommand{\ee}{\end{equation}}

\newcommand{\mincir}{\raise
-3.truept\hbox{\rlap{\hbox{$\sim$}}\raise4.truept\hbox{$<$}\ }}
\newcommand{\magcir}{\raise
-3.truept\hbox{\rlap{\hbox{$\sim$}}\raise4.truept\hbox{$>$}\ }}

\providecommand{\U}[1]{\protect\rule{.1in}{.1in}}

\usepackage{tikz,xcolor,hyperref}

\definecolor{lime}{HTML}{A6CE39}
\DeclareRobustCommand{\orcidicon}{%
	\begin{tikzpicture}
	\draw[lime, fill=lime] (0,0) 
	circle [radius=0.16] 
	node[white] {{\fontfamily{qag}\selectfont \tiny ID}};
	\draw[white, fill=white] (-0.0625,0.095) 
	circle [radius=0.007];
	\end{tikzpicture}
	\hspace{-2mm}
}

\foreach \x in {A, ..., Z}{%
	\expandafter\xdef\csname orcid\x\endcsname{\noexpand\href{https://orcid.org/\csname orcidauthor\x\endcsname}{\noexpand\orcidicon}}
}

\begin{document}

\title{\Large{Phase space analysis of the bouncing universe with stringy effects}}

\author{Alfredo D. Millano\orcidB{}$^1$}
\email{alfredo.millano@alumnos.ucn.cl}
\author{Kimet Jusufi\orcidA{}$^2$}
\email{kimet.jusufi@unite.edu.mk}
\author{Genly Leon\orcidC{}$^{1,3}$}
\email{genly.leon@ucn.cl}
\affiliation{$^1$Departamento de Matem\'{a}ticas, Universidad Cat\'{o}lica del Norte, Avda.
Angamos 0610, Casilla 1280 Antofagasta, Chile}
\affiliation{$^2$Physics Department, State University of Tetovo, 
Ilinden Street nn, 1200, Tetovo, North Macedonia}
\affiliation{$^3$Institute of Systems Science, Durban University of Technology, PO Box 1334, Durban 4000, South Africa}

\begin{abstract}
We use the recently modified Friedmann equations obtained from string T-duality effects that encode the zero-point length (\href{https://doi.org/10.1016/j.physletb.2022.137621}{Phys. Lett. B 836 (2023), 137621}) to study the phase space analyses of a bouncing early universe. An important implication of such stringy effects is that they can alleviate the initial singularity since the Raychaudhuri equation is modified. We investigate if the Universe can undergo an accelerated expansion phase for a specific domain of the equation of state parameter. The stringy effects are encoded in the parameter $\Gamma$, which depends not only on the zero-point length but also on the state parameter $\omega$. We construct two dynamical systems depending on whether $-1 < \omega \leq -1/3$ and $\omega \geq -1/3$, and we classify the equilibrium points of each system. Exact solutions and cosmological implications are discussed. 
\end{abstract}
\maketitle
\section{Introduction}
Although the tremendous success of the cosmological $\Lambda$CDM model in describing the current state and evolution of the Universe, the deep origin of the accelerating expansion of the Universe still needs to be well understood \cite{SupernovaSearchTeam:1998fmf, SupernovaCosmologyProject:1997zqe, SupernovaCosmologyProject:1998vns}. We often refer to the cosmological constant as dark energy - a hypothetical source of energy permeating the Universe and exerting a negative pressure, causing, in this way, the expansion of the Universe to accelerate \cite{Carroll:1991mt}. Nevertheless, dark matter is another mysterious form of matter revealed by different observations. There are a lot of observational facts that support the presence of dark matter and dark energy. It is well known that dark matter is needed to explain the galaxy rotation curves. In particular, one can use the rotation of stars in galaxies to infer the presence of matter/dark matter in a given galaxy. In this direction, the astrophysical measurements of the rotation of stars have revealed the presence of dark matter \cite{Salucci:2018eie}. 
Another essential fact is cosmic microwave background (CMB) radiation. The CMB is thought to be remnant radiation of the early Universe; it is characterised by highly uniform temperature, but, importantly, there are slight variations in temperature, which can be used to infer many properties of the early universe \cite{Mather:1990tfx, Mather:1998gm, WMAP:2003ivt, Planck:2013lks}. 
Let us mention here the observational fact related to the Type Ia supernovae. We can use this explosive death of stars to measure the distance to distant galaxies and discover that the Universe is expanding. That, again, implies the role of the cosmological constant in explaining the expansion of the Universe. Among other things, dark matter and dark energy are needed to explain the Universe's large-scale structure. Other findings in modern cosmology, like the baryon acoustic oscillations, have revealed further arguments supporting the existence of dark matter and dark energy \cite{Beutler:2011hx}. The variations in the density of matter in the early Universe can infer the properties of the Universe. 
Finally, we can use general relativistic effects like the gravitational lensing effect to infer the distribution of matter in large-scale structures. 
The $\Lambda$CDM model fits very well with the above observational facts. 

Other attempts to explain the Universe's accelerated expansion in terms of the quintessence field have been proposed \cite{Ratra:1987rm, Peebles:1987ek, Coble:1996te, Turner:1997npq, Caldwell:1997ii}. Unlike the cosmological constant, which is assumed to be a constant amount of energy density everywhere, quintessence is a dynamic field and, therefore, can vary with time. Another possibility to explain the acceleration rate of the Universe might be the modified gravity theories (see \cite{Dutta:2016ukw,Saridakis:2016ahq,Sheikhahmadi:2016wyz,Garcia-Salcedo:2000ujn,Camara:2004ap,Elizalde:2003ku,Quercellini:2008vh,Cai:2009zp,Saridakis:2010mf,Saridakis:2009bv,Leon:2009rc,Breton:2012yt,Xu:2012jf,Leon:2013qh,Kofinas:2014aka,Leon:2012vt,Fadragas:2013ina,Leon:2015via,Pulgar:2014cba,Leon:2014yua,Giacomini:2017yuk,Joyce:2016vqv,Tsujikawa:2013fta,Nojiri:2017ncd,Lepe:2017yvs,Harko:2014aja, Otalora:2013dsa,Otalora:2013tba,Karpathopoulos:2017arc,Leon:2012mt,Akarsu:2012dj,Akarsu:2014dxa,Akarsu:2011zd,Grasso:2000wj,Garrison:2016sty,Finke:2015ona,Haba:2016swv,Stachowski:2016dfi,Hrycyna:2014cka,Szydlowski:2012zz,Szydlowski:2008in,Dabrowski:2003jm,Szydlowski:2008by,Szydlowski:2006ay,Godlowski:2005tw,Szydlowski:2005ph,Garcia-Salcedo:2013cxa,Barrow:1990vx,Barrow:1990td,Barrow:2007zr,delCampo:2008dw,Cid:2011zbe,Herrera:2013ust,Cid:2015pja,Cid:2012kw,kierdorf,Jimenez:2001gg,Moresco:2012jh,Magana:2017nfs} and references therein). 

According to general relativity, our Universe started with the Big Bang, a cosmological singularity that can be understood in the Raychaudhuri equation. If we study the evolution of the Universe, one can show that the expansion scalar can become infinite at some point, indicating in this way the presence of the singularity. However, we now believe that such a singularity can be avoided in these extreme conditions if we consider quantum effects. 

In the 1970s, Hawking and Bekenstein showed a profound connection between black hole physics and the law of thermodynamics. In particular, it was argued that thermodynamic quantities such as entropy and temperature are proportional to the horizon area and surface
gravity, which are pure geometrical quantities. An exciting finding was achieved by Jacobson, who derived the Einstein field equation of gravity as the equation of state for spacetime \cite{Jacobson:1995ab}. This idea was further used in several studies; see in particular (\cite{Akbar:2006kj, Cai:2006rs, Cai:2005ra, Wang:2001bv, Wang:2001bf, Cai:2002ub, Cai:2008ys, Nojiri:2005pu, Sheykhi:2007zp,Sheykhi:2007gi, Sheykhi:2009zza,Sheykhi:2008qs, Sheykhi:2010zz,Sheykhi:2009zv}). Recently, Verlinde argued that gravity is not a
fundamental force but can be regarded as an entropic force
\cite{Verlinde:2010hp}. In Verlinde's gravity, gravity is an emergent effect, and dark matter is an apparent effect; an entropy displacement caused by matter \cite{Verlinde:2016toy}. Verlinde's argument is based on two important principles. The first principle is the
equipartition law of energy for the system's degrees of freedom, and the second is the holographic principle. In this line of thought, Padmanabhan made further steps to understand the nature of gravity. He argued that the spatial expansion of our Universe could be understood as
a consequence of the emergence of space. By equating the difference
between the number of degrees of freedom in bulk and on the
boundary with the volume change, he was able to obtain the Friedmann
equation describing the evolution of the Universe \cite{Padmanabhan:2009kr, Padmanabhan:2012ik} (see also \cite{Cai:2012ip, Yang:2012wn, Sheykhi:2013oac}).

Padmanabhan introduced the concept of duality and zero point length as a tool to obtain quantum gravity effect \cite{Padmanabhan:1996ap} and was applied to black hole physics to cure the black hole singularity \cite{Nicolini:2019irw, Gaete:2022ukm, Nicolini:2022rlz, Jusufi:2023pzt}. In the present work, we will study the phase space analyses of the recently modified Friedmann's equation found in \cite{Jusufi:2022mir}. The modified Friedmann's was obtained from Verlinde's emergent gravity with zero-point length corrections. The quantum effect is essential in the early Universe; in particular, it was shown how the singularity could be avoided, and the Universe can perform a bouncing scenario \cite{Jusufi:2022mir}. 

This paper is structured as follows. In Sec. \ref{II}, we review the modified Friedmann's equation in string T-duality derived from Verlinde's entropic force scenario. In Sec. \ref{III}, we investigate the phase space analyses of the early Universe. Finally, we comment on our findings in Sec. \ref{IV}. 
\section{Modified Friedmann equations with stringy effects}
\label{II}
First, we review the modified Fredmann's equation reported in \cite{Jusufi:2022mir}. The gravitational potential is modified, and, consequently, Newton's law reads \cite{Jusufi:2022mir}
\begin{equation}\label{F5}
F=-\frac{GMm}{R^2}\left[1+\frac{l_0^2}{R^2}\right]^{-3/2},
\end{equation}
here $l_0$ is the zero-point length with a value close to the
Planck length (see, \cite{Nicolini:2019irw}). We shall assume a spatially homogeneous and isotropic background described by the Friedmann-Robertson-Walker (FRW) metric
\begin{equation}
ds^2={h}_{\mu \nu}dx^{\mu} dx^{\nu}+R^2(d\theta^2+\sin^2\theta
d\phi^2),
\end{equation}
in which $R=a(t)r$, $x^0=t, x^1=r$. For the metric functions, we have
\begin{equation}
    h_{\mu \nu}=\diag\left(-1, \frac{a^2}{1-kr^2}\right),
\end{equation}
with $k$ being the
the curvature of space; in particular, for $k = 1, 0, -1$, we have a closed, flat, and open FLRW model, respectively. For the FRW metric, we can find the dynamical apparent
horizon, by the following equation
\begin{eqnarray}
h^{\mu
\nu}\partial_{\mu}R\partial_{\nu}R=0. 
\end{eqnarray}
It is easy to show that the apparent horizon radius for the FRW Universe reads
\begin{equation}
\label{radius}
 R=ar=\frac{1}{\sqrt{H^2+k/a^2}}.
\end{equation}
In addition, for the matter source in the FRW Universe, we shall assume a perfect fluid described in terms of the stress-energy tensor as follows
\begin{equation}\label{T}
T_{\mu\nu}=(\rho+p)u_{\mu}u_{\nu}+pg_{\mu\nu},
\end{equation}
along with the continuity equation
\begin{equation}\label{Cont}
\dot{\rho}+3H(\rho+p)=0.
\end{equation}
The last equation $H=\dot{a}/a$ defines the Hubble parameter. First, we can derive the dynamical equation and, for this goal, let us consider a compact spatial region $V$ with a compact boundary
$\mathcal S$, namely we have a sphere with a radius $R= a(t)r$, where $r$ is a dimensionless quantity. By combining the second law of Newton with the gravitational force (\ref{F5}),
we get
\begin{equation}\label{F6}
F=m\ddot{R}=m\ddot{a}r=-\frac{GMm}{R^2}\left[1+\frac{l_0^2}{R^2}\right]^{-3/2}.
\end{equation}
Next, to derive the Friedmann equations of the FRW Universe, we need to use $\mathcal M$ rather than the total mass $M$. By replacing $M$ with $\mathcal M$, it follows \cite{Jusufi:2022mir}
\begin{equation}\label{M1}
\mathcal M =-\frac{\ddot{a}
a^2}{G}r^3\left[1+\frac{l_0^2}{R^2}\right]^{3/2}
\end{equation}
and for the gravitational mass, we have
\begin{equation}\label{Int}
\mathcal M =2
\int_V{dV\left(T_{\mu\nu}-\frac{1}{2}Tg_{\mu\nu}\right)u^{\mu}u^{\nu}}.
\end{equation}
From these equations, it is easy to obtain the modified dynamical
evolution of the  FRW universe \cite{Jusufi:2022mir}
\begin{equation}\label{addot}
\frac{\ddot{a}}{a} =-\frac{4\pi
G}{3}(\rho+3p)\left[1+\frac{l_0^2}{R^2}\right]^{-3/2}.
\end{equation}

Furthermore,  since $l_0$ is a very small number, we can consider a series expansion around $l_0$ \cite{Jusufi:2022mir} 
\begin{equation}\label{addot0}
\frac{\ddot{a}}{a}=- \left(\frac{4 \pi G  }{3}\right)(\rho+3p)\left[1-\frac{3}{2} \frac{l_0^2}{r^2 a^2}+... \right],
\end{equation}
then by multiplying $2\dot{a}a$ on both sides of Eq. (\ref{addot}), and making use of the continuity equation 
\begin{eqnarray}
\rho=\rho_0 a^{-3 (1+w)},\label{(eq13)}
\end{eqnarray}
it was found that the modified Friedmann's and Raychaudhuri's equations \cite{Jusufi:2022mir}
\begin{equation}\label{Fried2}
H^2+\frac{k}{a^2} =\frac{8\pi G}{3}\rho\left[1-\Gamma \rho\right],
\end{equation}
and
\begin{equation}
\dot{H} - \frac{k}{a^2}=-4 \pi  G (\rho + p)  (1- 2 \Gamma \rho), \label{Ray2} 
\end{equation}
respectively, with $p=\omega \rho$ and $\Gamma$ defined by
\begin{equation}\label{Gamma}
\Gamma\equiv\frac{4\pi G \,l_0^2 }{3}\left(\frac{1+3
\omega}{1+\omega}\right).
\end{equation}

For $\Gamma>0$, one can discover that the string T-duality modified Friedmann equations have a similar form with the modified Friedmann's equation found in Loop quantum gravity (LQG)  \cite{Ashtekar:2011rm}
\begin{equation}
\left( \frac{\dot{a}}{a}\right)^2+\frac{k}{a^2} =\frac{8\pi G \rho }{3}\left[1-\frac{\rho}{\rho_c}\right],
\end{equation}
where $\rho_c$ is the critical energy density defined by
\begin{equation}
\rho_c\equiv \frac{3}{8 \pi \gamma^2 \lambda^2 },
\end{equation}
with $\lambda \sim 5.2 l_{Pl}^2$ \cite{Ashtekar:2011rm} being the area gap that sets the discreteness scale of Loop quantum gravity, and $\gamma$ is the Immirzi parameter. The correspondence is achieved by identifying $\Gamma=\rho_c^{-1}$ \cite{Jusufi:2023pzt}. One difference, however, is that the  $\Gamma$ depends not only on the zero-point length but also on the state parameter $\omega$. An interesting result can be found if we set $\dot{a}=0$ ($H=0$); in that case, we can solve the equation for the critical density we get \cite{Jusufi:2022mir} 
\begin{eqnarray}
\rho_{\rm {c}}=\frac{1}{2 \Gamma}\left(1 \pm  \frac{\sqrt{2 G \pi (2 G \pi a^2 -2 \Gamma  k)} }{2 G \pi a}  \right). \label{eq19}
\end{eqnarray}
The last equation imposes the constraint $2 G \pi a^2 -2 \Gamma k \geq 0$. Solving for the scale factor $a=a_{\rm min}$, we get 
\begin{eqnarray}
a_{\rm min}=\sqrt{2}\, l_0  \sqrt{\frac{(1+3 \omega)\,k}{1+\omega}}.
\end{eqnarray}
This equation shows that for a closed universe, there exists a minimal length for the scale factor and, in particular, for  $\omega=0$, we get $a_{\rm {min}}=\sqrt{2} \,l_0$. For a flat universe with $k=0$, the condition $\dot{a}=0$ ($H=0$) gives $\rho_{\rm c}=1/\Gamma$.

On the other hand, when $\Gamma<0 $, the modified Friedmann equation corresponds to the one of Brane-cosmology \cite{Randall:1999ee, Binetruy:1999hy, Binetruy:1999ut, Bowcock:2000cq, Maartens:2003tw, Brax:2004xh, Clifton:2011jh}, particularly, on the RS2 scenario with a single $Z_2$
symmetric brane \cite{Randall:1999ee}. Eqs.  (656) and (657) in \cite{Clifton:2011jh} are
\begin{align}
    H^2 +\frac{k}{a^2} &= \Lambda_4 + \frac{8 \pi G_4}{3} \rho \left(1+ \frac{\rho}{2 \sigma}\right) + \frac{\mu}{a^4},\\
    \dot{H } -\frac{k}{a^2} & = - 4 \pi G (\rho + p)\left(1+ \frac{\rho}{\sigma}\right) - 2 \frac{\mu}{a^4},
\end{align}
\noindent
where $H$ the Hubble parameter along the brane, $a$ is the scale factor, and
$k = 0, \pm 1$ describes the spatial curvature. A tension, $\sigma$ source the brane, and
a cosmological fluid with energy density, $\rho$, and pressure, $p=\omega \rho$. The parameters  $\Lambda_4$
and $G_4$ denote the effective cosmological constant and Newton's constant on the brane,
respectively. 

To compare with equations \eqref{Fried2} and \eqref{Ray2}, we set the dark radiation term to zero and the 4-dimensional cosmological constant to zero $\Lambda_4=0$. The brane tension is reinterpreted as $\sigma= {1}/{(2 |\Gamma|)}$. 

An interesting result can be found if we set $\dot{a}=0$ ($H=0$); in that case, we can solve the equation for the brane-tension to get 
\begin{eqnarray}
\sigma_{\rm c}=-\frac{1}{4\Gamma}\left(1 \pm  \frac{\sqrt{2 G \pi (2 G \pi a^2 -2 \Gamma  k)} }{2 G \pi a}  \right). \label{Neq19}
\end{eqnarray}
The last equation imposes the constraint $2 G \pi a^2 -2 \Gamma k \geq 0$. Solving for the scale factor $a=a_{\rm min}$, we get 
\begin{eqnarray}
a_{\rm min}=\sqrt{2}\, l_0  \sqrt{\frac{(1+3 \omega)\,k}{1+\omega}}, \quad \sigma|_{a_{\rm min}}= -\frac{1}{4\Gamma}. 
\end{eqnarray}
This equation shows that for a closed universe, there exists a minimal length for the scale factor and, in particular, for  $\omega=0$, we get $a_{\rm {min}}=\sqrt{2} \,l_0$. For a flat universe with $k=0$, the condition $\dot{a}=0$ ($H=0$) gives $\sigma= {1}/{(2 |\Gamma|)}$. 

Following \eqref{Neq19} for an open Universe with $k=-1$, we can get the critical brane-tension provided $\Gamma \leq 0$. Since $l_0$ is of Planck length order, i.e., $l_0 \sim l_{Pl}$, its effect becomes important only in short distances. The critical density that corresponds to the minimal scale goes like $\sigma_{\rm c} \sim l_{Pl}^{2}$, and it may prove a bound for the maximum brane-tension in nature.

In the present work, we consider the interval $\omega\in[-1,1]$ as the physical one. According to \eqref{Gamma}, we have two regimes 
\begin{equation}
\left\{ \begin{array}{ccc}
    \Gamma \leq 0,   &  -1<\omega \leq -\frac{1}{3}, & \text{Brane-like scenario}\\
    \Gamma \geq 0,  & \omega \geq -\frac{1}{3}, & \text{LQG-like scenario}
 \end{array}  \right.. 
\end{equation}

We see that in the limit $l_0 \to 0$, the standard Friedmann equation is obtained. Furthermore, we can derive the modified
Raychaudhuri equation using
\begin{eqnarray}
\dot{H}=-H^2+\frac{\ddot{a}}{a}.
\end{eqnarray}
and the deceleration parameter is defined as
\begin{equation}
q=-1-\frac{\dot{H}}{H^2},
\end{equation}
where depending on the sign of $q$ we can have deceleration or
acceleration scenario, respectively. As shown in \cite{Jusufi:2023pzt}, the modified Friedmman's equations have interesting implications. Take for simplicity a flat universe with $k=0$ along with the definition $H_0^2=8 \pi G \rho/3$ and can write \cite{Jusufi:2023pzt}
\begin{eqnarray}
H \simeq H_0 \left(1-\Delta_{GUP} H_0^2\right),
\end{eqnarray}
which has the form of Generalized Uncertainty Principle (GUP) modified Hubble parameter with $\Delta_{GUP}=\alpha l_0^2/2$ \cite{Maggiore:1993rv, Hossenfelder:2012jw, Aghababaei:2021gxe, Giacomini:2020zmv, Paliathanasis:2022xvn}. 
Since we believe that Cosmic Microwave Background  (CMB) carries quantum gravity fingerprints, this equation can have implications for the Hubble tension problem. For $H$, we can take the value reported by the Planck collaboration, $H=67.40 \pm 0.50$ km s$^{-1}$ Mpc$^{-1}$ \cite{Planck:2018vyg}. On the other hand, $H_0$ can be viewed as the unmodified Hubble parameter, and we can take the value reported by the Hubble Space Telescope (HST), $H_0=74.03 \pm 1.42$ km s$^{-1}$ Mpc$^{-1}$ \cite{Riess:2019cxk}, see also \cite{Dainotti:2021pqg, Dainotti:2022bzg}. 

\section{Phase space analysis}
\label{III}

 Now we present the evolution equations
\begin{align}
  \dot{H} &=\frac{k}{a ^2}+4 \pi  G (\omega +1) \rho  (2 \Gamma \rho -1), \label{evolH}\\
  \dot{\rho}& = -3 H(1+\omega) \rho. \label{evolrho}
\end{align}
as an autonomous dynamical system. We construct two dynamical systems depending on whether $-1 < \omega \leq -1/3$ and $\omega \geq -1/3$, and we classify the equilibrium points of each system, and cosmological implications are discussed. 

Firstly, we find exact solutions and find qualitative results. Exact solutions can play an important role in the evolution of whole classes of models acting as asymptotic or intermediate stages in the evolution of the Universe. Additionally, one can use qualitative techniques of dynamical systems to obtain relevant information about the flow properties associated with an autonomous system of ordinary differential equations. For example, the qualitative theory of differential equations \cite{wiggins, perko, Hirsch} has some applications in cosmology \cite{TWE, coleybook, Coley:1999uh}.

Notice that, by substituting \eqref{(eq13)} and $H=\dot{a}/a$ in \eqref{evolH} we obtain 
\begin{align}
    -a  \ddot{a}+{\dot{a}}^2-4 \pi  \rho_0 G (\omega +1) a ^{-6 \omega -4} \left(a^{3 \omega +3}-2 \Gamma  \rho_0\right)+k=0,
\end{align}
which admits the first integral 
\begin{align}
     t-t_U =\bigints _1^a {\left(\frac{8}{3} G \pi  \rho_0 \left(s^{3 \omega +3}-\Gamma  \rho_0\right) s^{-6 \omega -4}+c_1 s^2-k\right)^{-\frac{1}{2}}}ds
\end{align}
where we have denoted $t_U$ the Universe's age, and we set $a(t_U)=1$ for the current value of the scale factor and $\rho_0$ for the current value of the energy density. The integration constant $c_1$ satisfies 
\begin{equation}
    c_1= \dot{a}^2(t_U)+\frac{8}{3} \pi  G  \rho_0 (\Gamma  \rho_0 -1)+k. 
\end{equation} Therefore, the system is integrable.

Now we are interested in the asymptotic behaviour of the solutions. We consider two separate cases according to the sign of $\Gamma$.

\subsection{$\Gamma\leq 0$}

This regime corresponds to $-1<\omega \leq -\frac{1}{3}$. Then, the Friedmann equation \eqref{Fried2} becomes 

\begin{equation}
H^2+\frac{k}{a^2} =\frac{8\pi G}{3}\rho + \frac{8\pi G}{3} |\Gamma| \rho^2, \label{constr_1}
\end{equation}
where for the closed Universe, we set $k=+1$.

The quantities in the left-hand and right-hand sides of \eqref{constr_1} are non-negative. 
Then, we can define de dimensionless variables 
\begin{align}
  & \Omega=  \frac{8\pi G}{3 D^2}\rho, \; \Omega_k= \frac{k}{a^2 D^2}, \; Q=\frac{H}{D}, \;  D= \sqrt{H^2+\frac{k}{a^2}},
\end{align}
where $(\Omega, \Omega_k, Q) \in [0,1] \times [0,1] \times [-1,1]$, and satisfy 
\begin{align}
    & \frac{3 |\Gamma|  \Omega ^2 H^2}{8\pi  G}= (1-\Omega) (1-\Omega_{k}), \\
    & Q^2+\Omega_k=1. \label{Omega_k}
\end{align}
Introducing the time derivative 
\begin{equation}
    \frac{d f}{d\tau}= \frac{1}{\sqrt{H^2 +k a^{-2}}}\dot{f},
    \end{equation}
    we obtain the dynamical system 
\begin{align}
\frac{d \Omega_k}{d\tau} & =  Q \Omega_{k} (-3 \omega  (\Omega -2)-3 \Omega   +4),\\
\frac{d \Omega}{d\tau} & = -3 Q (\omega +1) (\Omega -1) \Omega,\\
\frac{d Q}{d\tau} & =   \frac{1}{2}  \Omega_{k} (3 \omega  (\Omega -2)+3 \Omega -4).
\end{align}
Using the equation \eqref{Omega_k} as a definition of $\Omega_k$ we obtain the reduced dynamical system 
\begin{align}
\frac{d \Omega}{d\tau} & = -3 Q (\omega +1) (\Omega -1) \Omega, \label{syst_1}\\
\frac{d Q}{d\tau} & =   \frac{1}{2}  (1-Q^2) (3 \omega  (\Omega -2)+3 \Omega -4). \label{syst_2}
\end{align}
defined on the phase plane 
\begin{equation}
\left\{(\Omega,  Q) \in [0,1] \times [-1,1]\right\}.
\end{equation}
The system system \eqref{syst_1}
and \eqref{syst_2} admits the first integral 
\begin{align}
    Q(\tau )^2-1=e^{-\frac{c_1}{3 (\omega +1)}} (1-\Omega (\tau ))^{-\frac{3 \omega +1}{3 (\omega +1)}} \Omega (\tau
   )^{\frac{2 (3 \omega +2)}{3 (\omega +1)}}, \label{first-integral_1}
\end{align}
where $c_1$ is an integration constant. 

Then, we obtain from \eqref{syst_1} the equation 
\begin{align}
    \Omega '(\tau )=-3 (\omega +1) \epsilon  (\Omega (\tau )-1) \Omega (\tau ) \sqrt{1+e^{-\frac{c_1}{3 \omega +3}}
   (1-\Omega (\tau ))^{\frac{2}{3 (\omega +1)}-1} \Omega (\tau )^{2-\frac{2}{3 (\omega +1)}}},
\end{align}
where $\epsilon = \pm 1$ is the sign of the initial value $H(0)$. Finally, we have the quadrature: 
\begin{align}
    \tau = \pm \frac{c_2-\bigints _1^{\Omega }\frac{1}{(\xi-1) \xi \sqrt{e^{-\frac{c_1}{3 \omega +3}}
   (1-\xi)^{\frac{2}{3 (\omega +1)}-1} \xi^{2-\frac{2}{3 (\omega +1)}}+1}}d\xi}{3 (\omega +1)},
\end{align}
where $c_2$ is an integration constant. 

Equation \eqref{first-integral_1} suggests to define the function 
\begin{align}
Z(\Omega, Q):= (1-\Omega)^{\frac{3 \omega +1}{3 (\omega +1)}} \Omega^{-\frac{2 (3 \omega +2)}{3 (\omega +1)}} \left[1-  Q^2\right],
\end{align}
which satisfies $Z'=0$, implies $Z(\Omega, Q)=c$ where $c$ is a constant. 
From this equation, it follows that asymptotic solutions satisfy $\Omega \rightarrow 1$, or $\Omega \rightarrow 0$ or $Q^2 \rightarrow 1$.

The equilibrium points of system \eqref{syst_1}
and \eqref{syst_2} for the parameter region $-1<\omega \leq -\frac{1}{3}$ are presented in Tab. \ref{tab1}

\begin{table*}
\begin{tabular}{|c|c|c|c|c|c|c|}
\hline
Label & $\Omega$ & $Q$ & Existence & $k_1$ & $k_2$ & Stability \\\hline 
$P_1 $ & $0$ & $-1$ & always &  $-3 (\omega +1)$ & $-2(3 \omega+2)$ & sink for \\
  &&&&&& $-\frac{2}{3}<w <-\frac{1}{3}$\\
  &&&&&& saddle for \\
  &&&&&& $-1<\omega <-\frac{2}{3}$ \\
$P_2 $ & $0$ & $1$ & always & $3 (\omega +1)$ &$ 2(3 \omega+2)$ & source for \\
  &&&&&& $-\frac{2}{3}<w <-\frac{1}{3}$\\
  &&&&&& saddle for \\
  &&&&&& $-1<\omega <-\frac{2}{3}$ \\
$P_3 $ & $1$ &$ -1$ & always & $-(3 \omega +1)$ & $3 (\omega +1)$ & source for \\
  &&&&&&  $-1<\omega <-\frac{1}{3}$ \\
$P_4 $ & $1$ &$ 1$ & always &$ 3 \omega +1$ & $-3 (\omega +1)$ & sink for  \\
  &&&&&& $-1<\omega <-\frac{1}{3}$ \\
$P_5 $ &$2-\frac{2}{3 (\omega +1)}$ &$ 0$ & $-\frac{2}{3}<\omega \leq-\frac{1}{3}$ &$ -\sqrt{-(3 \omega +1) (3 \omega
   +2)} $& $\sqrt{-(3 \omega +1) (3 \omega +2)}$ & saddle \\\hline 
\end{tabular}
\caption{ \label{tab1} Equilibrium points of system \eqref{syst_1}
and \eqref{syst_2} for the parameter region  $-1<\omega\leq -\frac{1}{3}$. }
\end{table*}

The equilibrium points are the following. 

$P_1$: $(\Omega, Q) =(0,-1)$, always exists. It corresponds to a flat contracting solution that is a  saddle for $-1<\omega <-\frac{2}{3}$ or a sink for $-\frac{2}{3}<\omega <-\frac{1}{3}$.

$P_2$: $(\Omega, Q) =(0, 1)$, always exists.  It corresponds to an expanding flat solution that is a saddle for $-1<\omega<-\frac{2}{3}$  or a source for $-\frac{2}{3}<\omega <-\frac{1}{3}$.

$P_3$: $(\Omega, Q) = (1, -1)$, always exists.  It corresponds to a flat contracting matter-dominated solution that is a source for  $-1<\omega <-\frac{1}{3}$, and it mimics a quintessence fluid.

$P_4$: $(\Omega, Q) = (1, 1) $, always exists.  It corresponds to a flat expanding matter-dominated solution that is a sink for  $-1<\omega <-\frac{1}{3}$, and it mimics a quintessence fluid.

$P_5$:   $(\Omega, Q) =\left(2-\frac{2}{3 (\omega +1)}, 0\right)$ exists for $-\frac{2}{3}<\omega \leq -\frac{1}{3}$. It corresponds to the Einstein static universe, and it is a saddle.

Finally, for $\omega=-\frac{2}{3}$ we have the line of equilibrium points $\Omega=0,$ $-1\leq Q \leq 1$ with eigenvalues $\{0,Q\}$ this line is stable for $Q<0$ and unstable for $Q>0.$ For $\omega=-\frac{1}{3}$ we have the line of equilibrium points $\Omega=1,$ $-1\leq Q \leq 1$ with eigenvalues $\{0,-2Q\}$ this line is stable for $Q>0$ and unstable for $Q<0.$ 

 \begin{figure*}[ht!]
      \centering
      \includegraphics[scale=0.7]{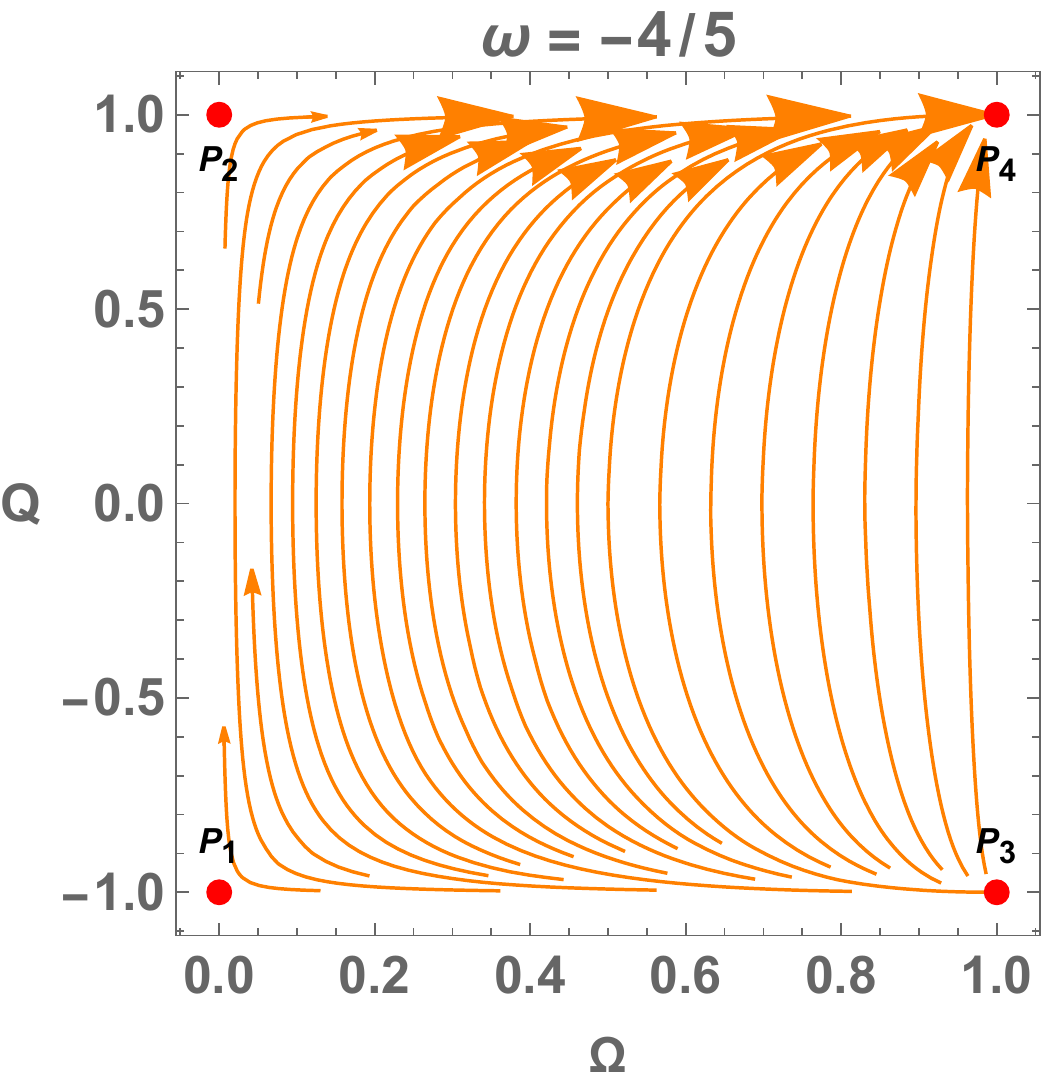}
      \includegraphics[scale=0.7]{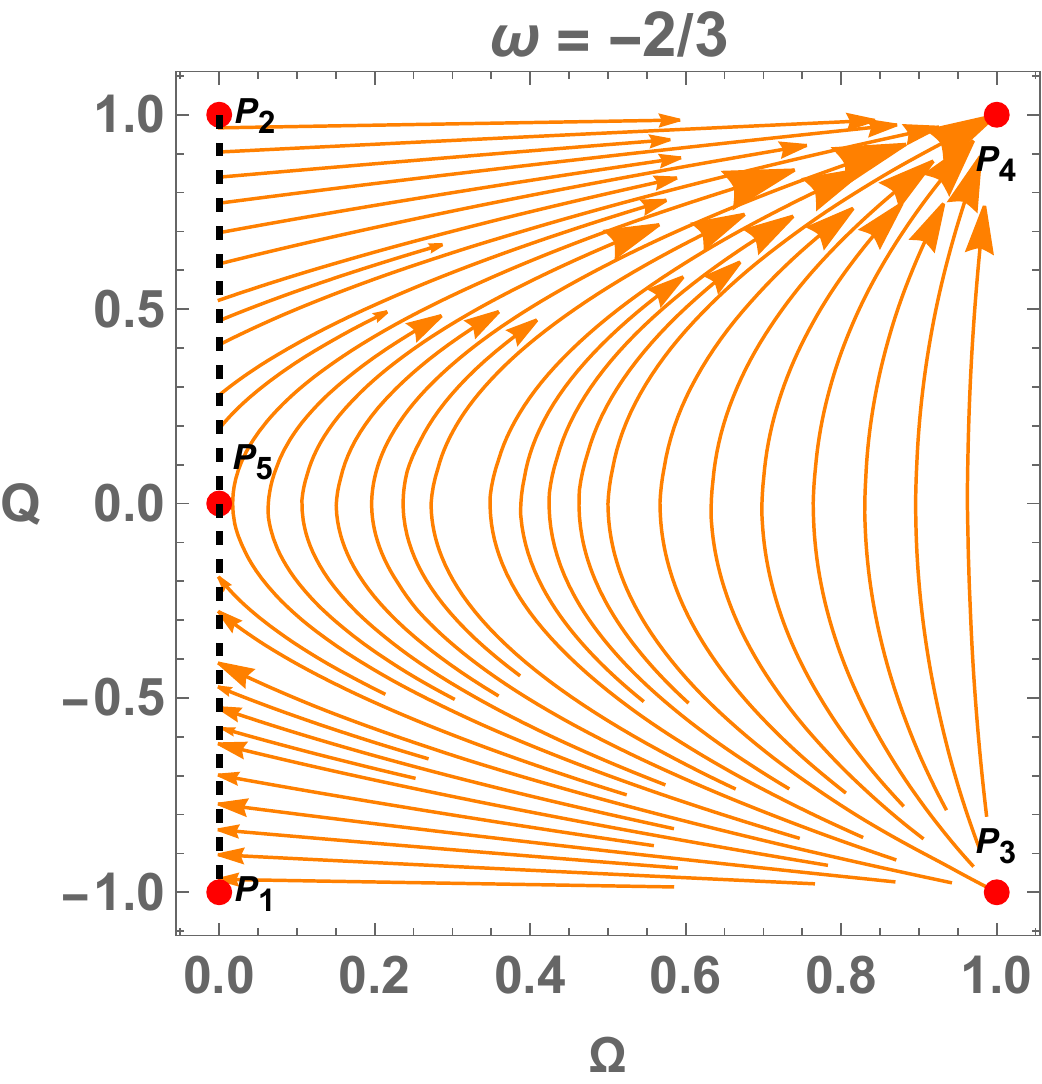}
      \includegraphics[scale=0.7]{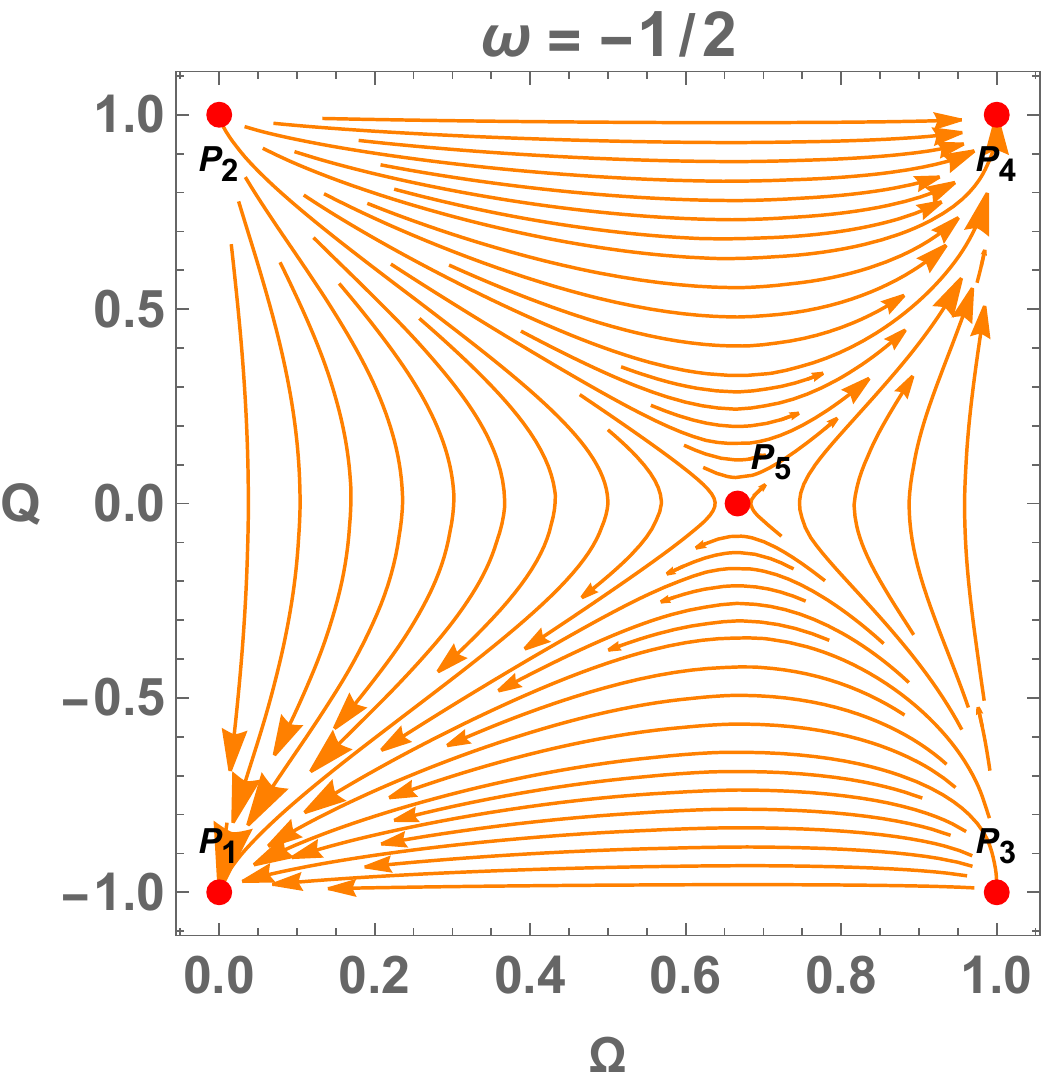}
      \includegraphics[scale=0.7]{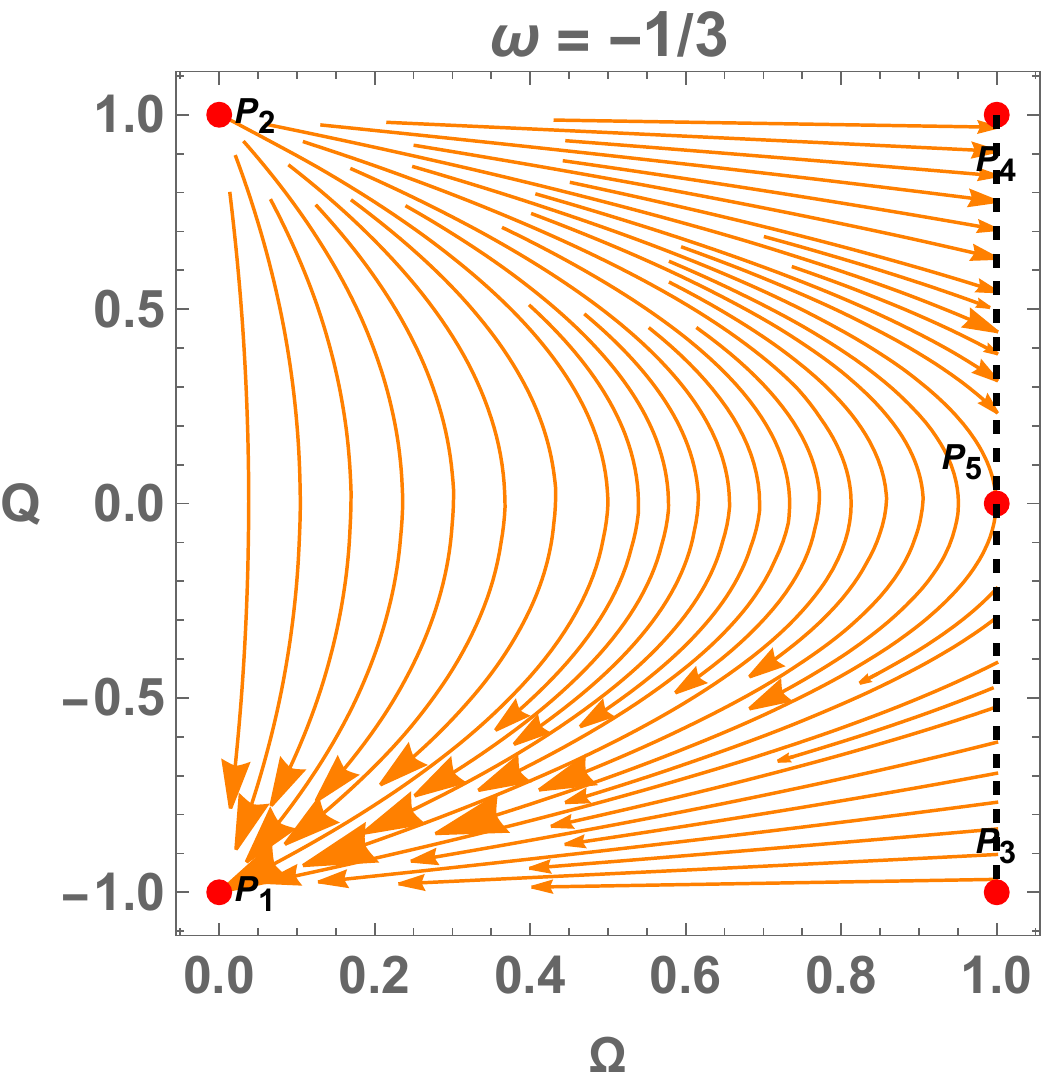}
      \caption{Phase plot for system \eqref{syst_1}, \eqref{syst_2} for different choices of the parameter $\omega.$}
      \label{fig:1}
  \end{figure*}

\subsection{$\Gamma\geq 0$}

This regime corresponds to $ \omega \geq -\frac{1}{3}$. Then, the Friedmann equation \eqref{Fried2} becomes 

\begin{equation}\label{constr_2}
H^2+\frac{k}{a^2} + \frac{8\pi G}{3}  \Gamma  \rho^2 =\frac{8\pi G}{3}\rho,
\end{equation}
where for the closed Universe, we set $k=+1$. 
The evolution equations are \eqref{evolH} and \eqref{evolrho}. 

The quantities in the left-hand and right-hand sides of \eqref{constr_2} are non-negative. 
Then, we can define de dimensionless variables 
\begin{align}
& {\tilde{Q}}=\frac{H}{\tilde{D}}, \;  \tilde{\Omega}=   \Gamma \rho, \; \tilde{\Omega}_k= \frac{k}{a^2 \tilde{D}^2}, \;  \tilde{D}= \sqrt{\frac{8\pi G}{3}\rho},
\end{align}
Constraint \eqref{constr_2} becomes 
\begin{equation}\label{constr_2b}
{\tilde{Q}}^2+\tilde{\Omega}_k + \tilde{\Omega} =1.
\end{equation}
Introducing the time derivative 
\begin{equation}
    \frac{d f}{d\eta}= {\tilde{D}}^{-1}\dot{f},
    \end{equation}
    we obtain the dynamical system 
    \begin{align}
       \frac{d {\tilde{Q}}}{d\eta}& = \frac{3}{2} (\omega +1) \left({\tilde{Q}}^2+2  \tilde{\Omega} -1\right)+ \tilde{\Omega}_k,\\
       \frac{d {\tilde{\Omega}_k}}{d\eta}& = {\tilde{Q}} (3 \omega +1)  \tilde{\Omega}_k,\\
       \frac{d {\tilde{\Omega}}}{d\eta}& = -3 {\tilde{Q}} (\omega +1)  \tilde{\Omega}. 
    \end{align}
    Using \eqref{constr_2b} as a definition of $\tilde{\Omega}$, we obtain the reduced system 
   \begin{align}
          \frac{d {\tilde{Q}}}{d\eta}& =   \tilde{\Omega}_k -\frac{3}{2} (\omega +1) \left({\tilde{Q}} ^2+2  \tilde{\Omega}_k -1\right), \label{syst_1b}\\
          \frac{d {\tilde{\Omega}_k}}{d\eta}& =  (3 \omega +1) {\tilde{Q}}  \tilde{\Omega}_k,  \label{syst_2b}
   \end{align} 
defined in the phase space 
  \begin{equation}
    \left\{(\tilde{Q}, \tilde{\Omega}_k ) \in [-1,1]\times [0, 1] :  {\tilde{Q}}^2+\tilde{\Omega}_k \leq 1\right\}. 
  \end{equation} 

The equilibrium points of system \eqref{syst_1b}
and \eqref{syst_2b} for the parameter region  $\omega\omega\geq -\frac{1}{3}$ are presented in Tab. \ref{tab2}.

\begin{table*}
\begin{tabular}{|c|c|c|c|c|c|c|}
\hline
Label & $\tilde{Q}$ & $\tilde{\Omega}_k$ & Existence & $k_1$ & $k_2$ & Stability \\\hline 
$Q_1$ &$ -1$ &$ 0$ &  always & $3 (\omega +1)$ & $-(3 \omega +1)$  & saddle for $\omega >-\frac{1}{3}$\\ \hline
$Q_2$ & $1$ &$ 0$ & always  & $-3 (\omega +1)$ &$ 3 \omega +1$   & saddle for $\omega >-\frac{1}{3}$\\ \hline
$Q_3$ & $0$ & $\frac{3 (\omega +1)}{6 \omega +4} $ & $\omega \geq -\frac{1}{3}$ & $-\sqrt{\frac{3}{2}} \sqrt{-((\omega+1) (3 \omega +1))} $& $\sqrt{\frac{3}{2}} \sqrt{-((\omega +1) (3\omega +1))}$ & centre  for $\omega> -\frac{1}{3}$ \\\hline
 \end{tabular}
\caption{ \label{tab2} Equilibrium points of system \eqref{syst_1b}
and \eqref{syst_2b} for the parameter region  $\omega\geq -\frac{1}{3}$. N.H. stands for nonhyperbolic}
\end{table*}

 \begin{figure*}[h]
      \centering
      \includegraphics[scale=0.7]{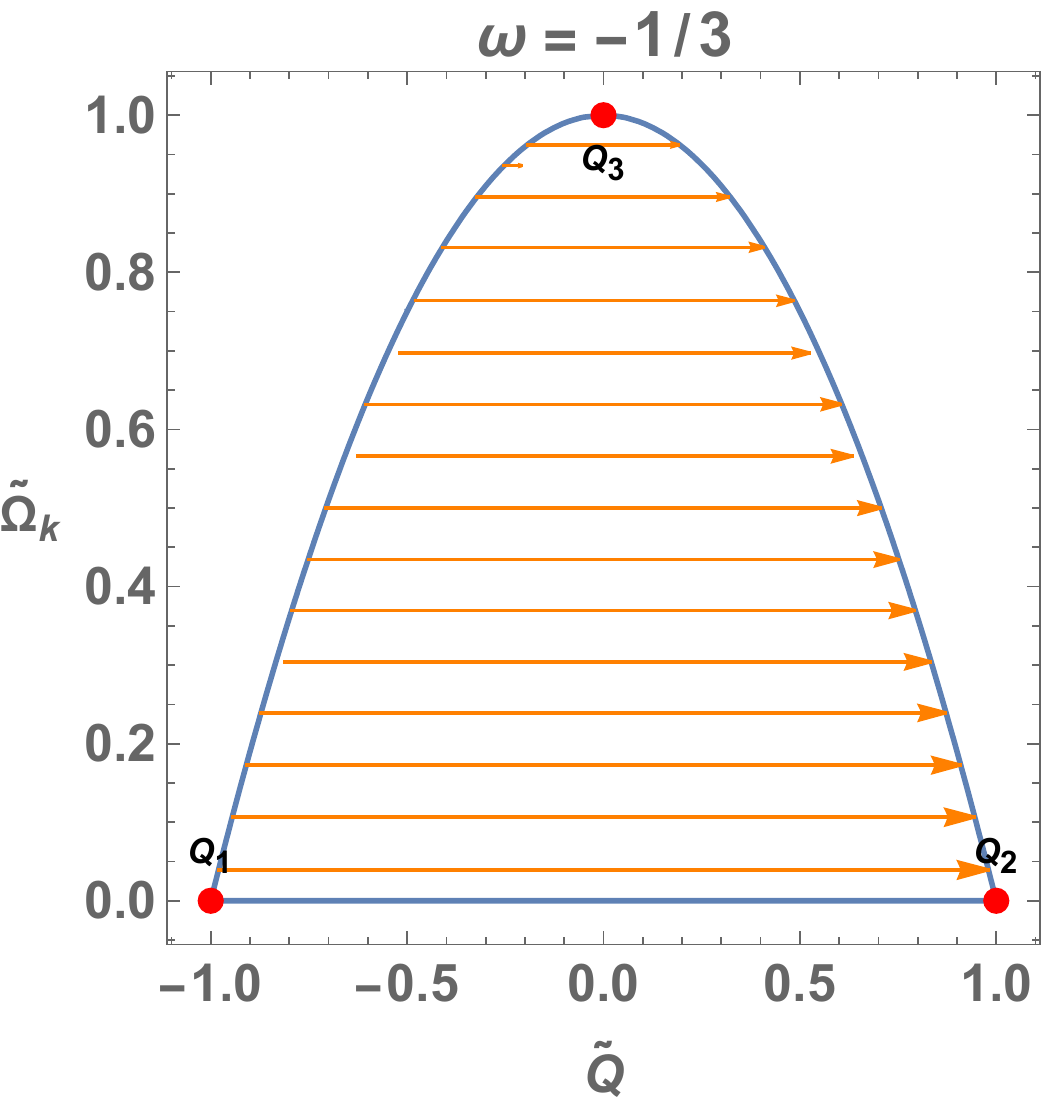}
      \includegraphics[scale=0.7]{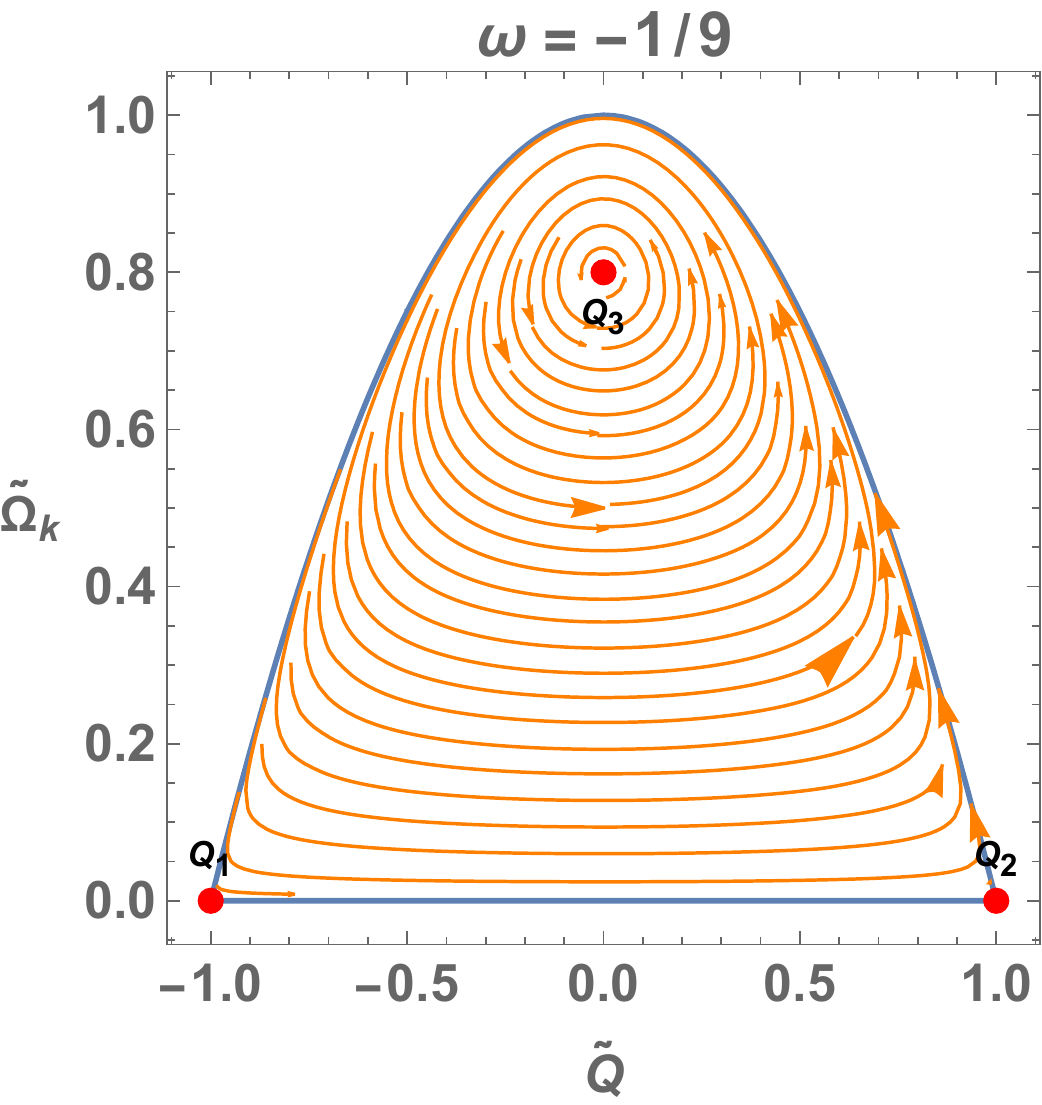}
      \includegraphics[scale=0.7]{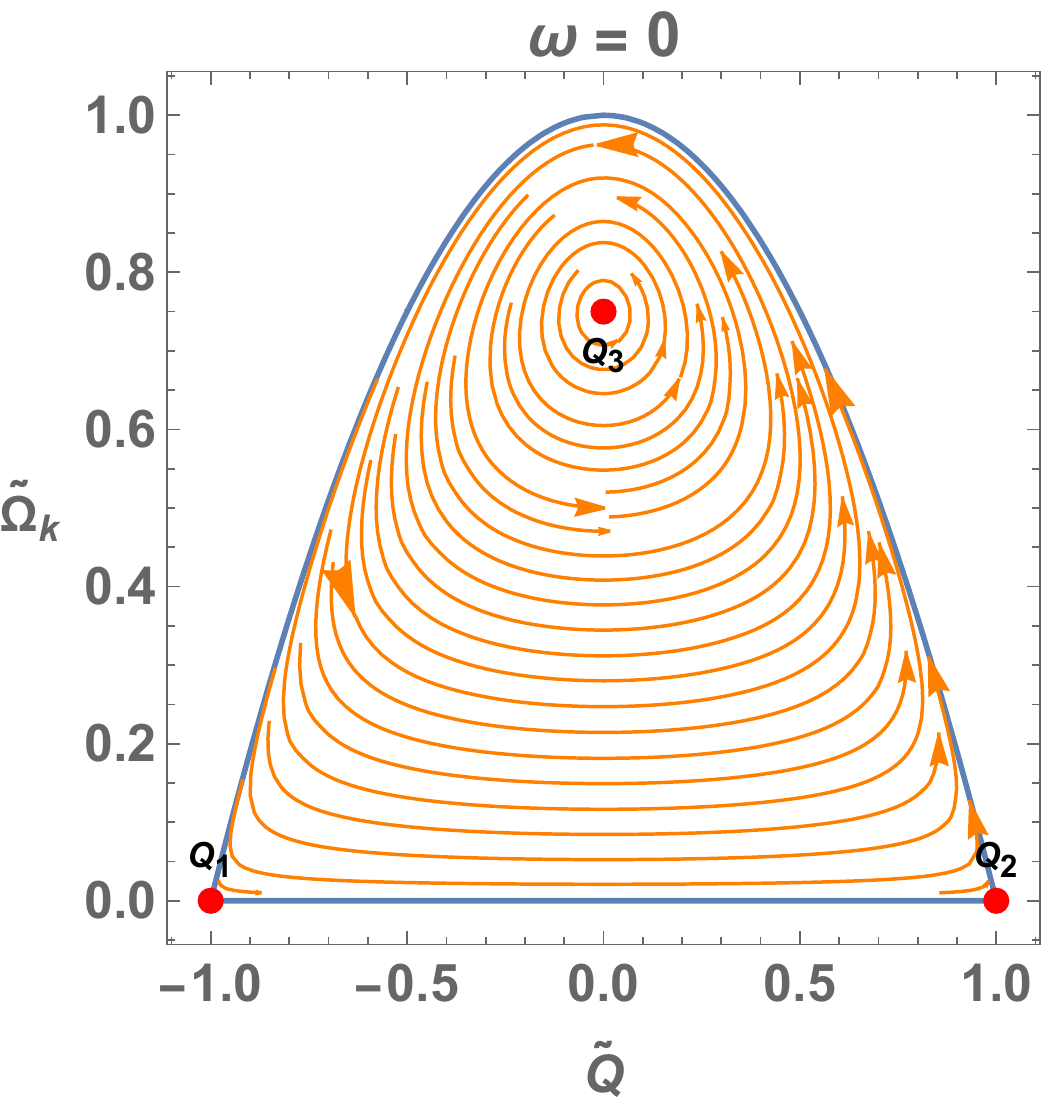}
      \includegraphics[scale=0.7]{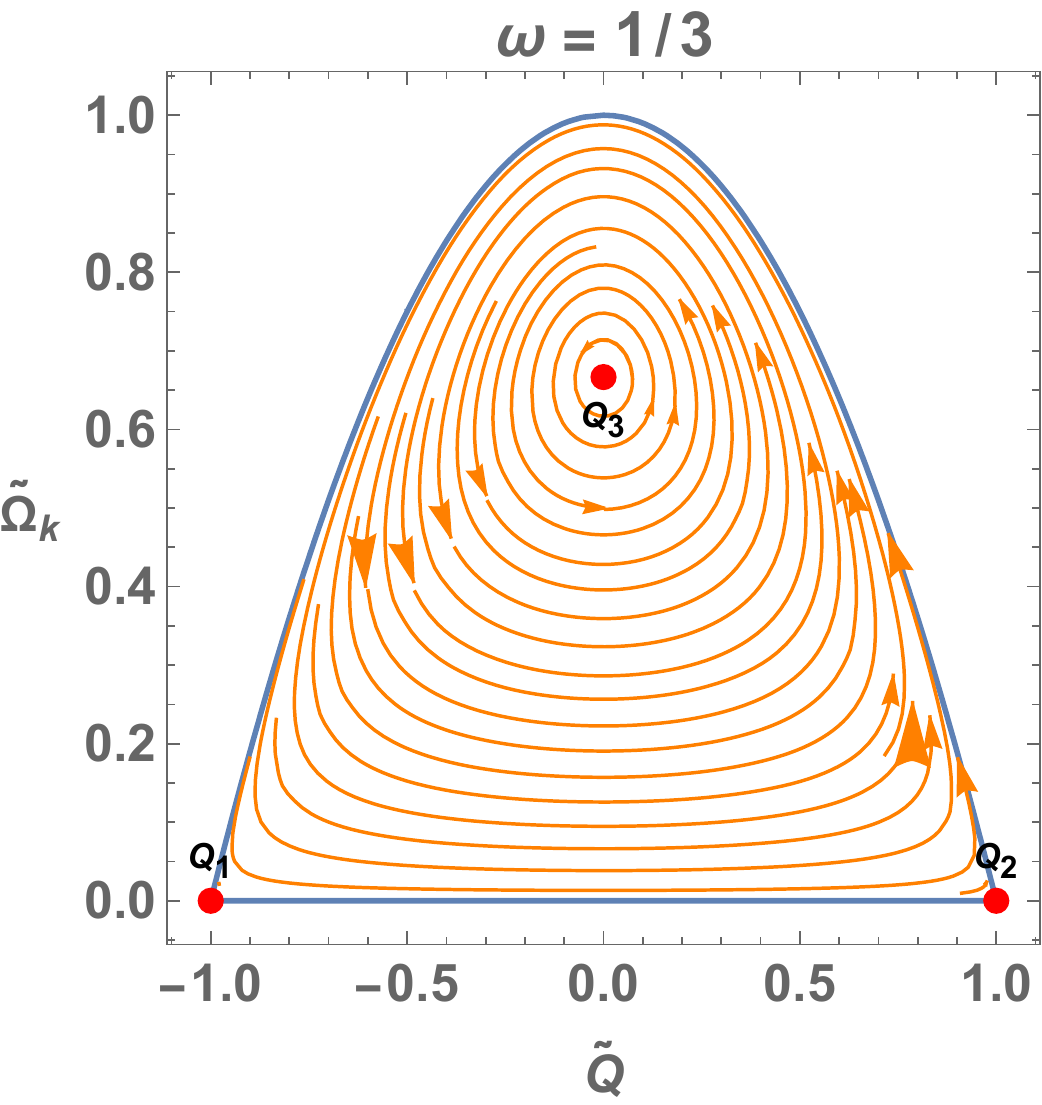}
      \caption{Phase plot for system \eqref{syst_1b}, \eqref{syst_2b} for different choices of the parameter $w.$}
      \label{fig:2}
  \end{figure*}

The equilibrium points for this case are the following.

$Q_1$: $\left(\tilde{Q}, \tilde{\Omega}_k\right)=(-1,0)$, always exists. It corresponds to a flat contracting solution. The eigenvalues are  $3 (\omega +1)$  and  $-(3 \omega +1)$, therefore, it is a saddle for $\omega >-\frac{1}{3}$.

$Q_2$: $\left(\tilde{Q}, \tilde{\Omega}_k\right)=(1,0)$, always exists.  It corresponds to a flat expanding solution. The eigenvalues are  $-3 (\omega +1)$ and $ 3 \omega +1$, therefore, it is a saddle for  $\omega >-\frac{1}{3}$.

$Q_3$: $\left(\tilde{Q}, \tilde{\Omega}_k\right)= \left(0,\frac{3 (\omega+1)}{6 \omega +4}\right)$, exists for $\omega \geq -\frac{1}{3}$. The eigenvalues are  $-\sqrt{\frac{3}{2}} \sqrt{-((\omega +1) (3 \omega +1))} $ and  $\sqrt{\frac{3}{2}} \sqrt{-((\omega +1) (3 \omega+1))}$. 
It corresponds to the Einstein static universe, a centre (nonhyperbolic) for $\omega> -\frac{1}{3}$. 

Finally, for $\omega=-1/3$ we obtain a line of equilibrium points $ {\tilde{Q}}^2+\tilde{\Omega}_k = 1$ with eigenvalues $\{0,-2Q\}$ that is unstable for $Q<0$ and stable for $Q>0$.

\section{Concluding Remarks} \label{IV}
In the present work, we provided detailed phase space analyses of the modified Friedmann's equations that describe the evolution of the early Universe. The modified Friedmann's equations are obtained from Verlinde's entropic and emergent force scenario and encode the string T-duality effect using the minimal length effect. One of the main implications of such modifications is the initial singularity of the early Universe. In particular, as was shown in \cite{Jusufi:2022mir}, the initial singularity is avoided due to the quantum gravity effects. That can be explained by the fact that there exists a minimal length scale $l_0$. 
An interesting result can be found if we set $\dot{a}=0$ ($H=0$); for $\Gamma>0$, one can find the contraction, or the collapsing matter stops at some critical distance $\sqrt{2} l_0$ and some critical density $\rho_{\rm c}$ (in the loop-quantum gravity scenario). Such a state is expected to be unstable and provides a natural way of generating an early expanding universe. However, in that case, for $\Gamma<0$, we can solve the equation for the brane-tension to get \eqref{Neq19}, that imposes the constraint $2 G \pi a^2 -2 \Gamma k \geq 0$. Solving for the scale factor $a=a_{\rm min}$, we get that for a closed universe, there exists a minimal length for the scale factor and, in particular, for  $\omega=0$, we get $a_{\rm {min}}=\sqrt{2} \,l_0$. For a flat universe with $k=0$, the condition $\dot{a}=0$ ($H=0$) gives $\sigma= {1}/{(2 |\Gamma|)}$. 

Following \eqref{Neq19} for an open Universe with $k=-1$, we can get the critical brane-tension provided $\Gamma < 0$. Since $l_0$ is of Planck length order, i.e., $l_0 \sim l_{Pl}$, its effect becomes important only in short distances. The critical density that corresponds to the minimal scale goes like $\sigma_{\rm c} \sim l_{Pl}^{2}$, and it may prove a bound for the maximum density in nature. 

Moreover,  we construct two dynamical systems depending on whether $-1 < \omega \leq -1/3$ and $\omega \geq -1/3$, and we classify the equilibrium points of each system. Finally, exact solutions and cosmological implications are discussed. 

Here we have first studied the phase space analyses for the case $\Gamma\leq 0$, or equivalently with the regime corresponding to $-1<\omega\leq -\frac{1}{3}$. In this regime, when $\Gamma<0 $, we can interpret the modified Friedmann equation as the one corresponding to the one of Brane-cosmology \cite{Randall:1999ee, Binetruy:1999hy, Binetruy:1999ut, Bowcock:2000cq, Maartens:2003tw, Brax:2004xh, Clifton:2011jh}, remarkably, on the RS2 scenario with a single $Z_2$ symmetric brane. Moreover, the matter field behaves as a quintessence fluid for the region of the equation of state presented. Our model mimics the effect of a scalar field trapped on an FLRW-RS2 braneworld \cite{Leyva:2009zz, Escobar:2011cz}.

Among other things, we found the following equilibrium points: $P_1$ that corresponds to a flat contracting solution that is a  saddle for $-1<\omega <-\frac{2}{3}$ or a sink for $-\frac{2}{3}<\omega <-\frac{1}{3}$; $P_2$, that corresponds to an expanding flat solution that is a saddle for $-1<\omega <-\frac{2}{3}$  or a source for $-\frac{2}{3}<\omega <-\frac{1}{3}$; $P_3$, that corresponds to a flat contracting matter-dominated solution that is a source for  $-1<\omega <-\frac{1}{3}$, and it mimics a quintessence fluid; $P_4$, that corresponds to a flat expanding matter-dominated solution that is a sink for  $-1<\omega <-\frac{1}{3}$, and it mimics a quintessence fluid and $P_5$ that exists for $-\frac{2}{3}<\omega \leq -\frac{1}{3}$ and corresponds to the Einstein static universe, and it is a saddle.

\bigskip

Then, we elaborated in more detail the domain  $\Gamma\geq 0$, or equivalently the region $ \omega \geq -\frac{1}{3}$ (Loop quantum gravity cosmology). In this case, we found the equilibrium points: $Q_1$, which corresponds to a flat contracting solution, and it is a saddle for $\omega >-\frac{1}{3}$; $Q_2$, that corresponds to a flat expanding solution and it is a saddle for  $\omega >-\frac{1}{3}$ and $Q_3$, which corresponds to the Einstein static universe and it is a centre (nonhyperbolic) for $\omega> -\frac{1}{3}$. The important finding is that the quantum gravity effects can provide a natural way of generating a dynamical early universe.

\acknowledgements{ADM was supported by Agencia Nacional de Investigación y Desarrollo (ANID) Subdirección de Capital Humano/Doctorado Nacional/año 2020 folio 21200837, Gastos operacionales Proyecto de tesis/2022 folio 242220121, and by Vicerrectoría de Investigación y Desarrollo Tecnológico (Vridt) at Universidad Católica del Norte. GL was funded through Concurso De Pasantías De Investigación Año 2022, Resolución Vridt No. 040/2022 and Resolución Vridt No. 054/2022. He also thanks the support of Núcleo de Investigación Geometría Diferencial y Aplicaciones, Resolución Vridt N°096/2022.}


\end{document}